\documentclass[twocolumn,aps,showpacs,floatfix,prc]{revtex4-2}
\usepackage{graphicx}
\usepackage[dvips]{epsfig}
\newcommand{\be}{\begin{equation}}    
\newcommand{\ee}{\end{equation}}
\newcommand{\beq}{\begin{eqnarray}}
\newcommand{\eeq}{\end{eqnarray}}
\newcommand{\beqn}{\begin{eqnarray*}}
\newcommand{\eeqn}{\end{eqnarray*}}
\def\lsim{\mathrel{\rlap{\lower2.5pt\hbox{\hskip1pt$\sim$}}
    \raise1pt\hbox{$<$}}}         
\def\gsim{\mathrel{\rlap{\lower2.5pt\hbox{\hskip1pt$\sim$}}
    \raise1pt\hbox{$>$}}}         

\def\nn{\nonumber}
\def\apr2{{\rm APR2~}}
\def\s1{{\rm BBS1~}}
\def\bs2{{\rm BBS2~}}
\def\g240{{\rm G240~}}
\def\ss1{{\rm SS1~}}
\def\sc2{{\rm SS2~}}
\def\lappreq{\mathrel{\rlap{\lower2.5pt\hbox{\hskip1pt$\sim$}}
    \raise1pt\hbox{$<$}}}         
\def\gappreq{\mathrel{\rlap{\lower2.5pt\hbox{\hskip1pt$\sim$}}
    \raise1pt\hbox{$>$}}}         
\begin{document}

\title{Universal relationships for neutron stars from perturbative approach}

\author{ Debasis Atta$^{1\dagger}$, Vinay Singh$^{2\S}$ and D. N. Basu$^{3\S}$ }

\affiliation{$^{\dagger}$Department of Higher Education, Government of West Bengal, Bikash Bhavan, Saltlake, Kolkata 700091, India}

\affiliation{$^{\S}$Variable  Energy  Cyclotron Centre, 1/AF Bidhan Nagar, Kolkata 700064, India}

\email[E-mail 1: ]{debasisa906@gmail.com}
\email[E-mail 3: ]{vsingh@vecc.gov.in}
\email[E-mail 3: ]{dnb@vecc.gov.in}

\date{\today }

\begin{abstract}

    The universal relationships for compact stars have been investigated employing perturbative approach using canonical (APR) and Brussels-Montreal Skyrme (BSk22, BSk24, BSk26) equations of state describing hadronic matter of neutron stars. The neutron star matter has been considered to be $\beta$-equilibrated neutron-proton-electron-muon matter at the core with a rigid crust. The multipole moments of a slowly rotating neutron star characterize its external gravitational field. These variables are dependent on the interior structure of the neutron star described by the equation of state of the neutron star matter. The properties of neutron stars, such as the mass, the radius, the dimensionless moment of inertia, the compactness, the Love number, the dimensionless tidal deformability and the dimensionless quadrupole moment have been calculated and relations among these quantities have been explored. It is found that most of these relations do not depend sensitively on the details of the internal structure of neutron stars. Such universality implies that the measurement of a single quantity appearing in a universal relation would automatically provide information about the others, notwithstanding the fact that those may not be accessible observationally. These can be used to estimate deformability of compact stars through moment of inertia measurements, to quantify spin in binary inspirals by breaking degeneracies in the detection of gravitational waves and test General Relativity in a way that is independent of nuclear structure.
		
\vspace{0.2cm}    

\end{abstract}


\maketitle

\noindent
\section{Introduction}
\label{sec:Section 1} 
    
    The compact stars provide the most conspicuous natural laboratories for exploring the matter constituting these objects \cite{Glendenning-1997,Haensel-2007,Bielich-2020,Weber-1999} at the extremes of density, pressure and isospin asymmetry. Gravitational waves (GWs) detected from recently found pulsars, like GW170817, have provided fresh insights into the internal structure of these stars, primarily regarding composition \cite{Abbott-1,Abbott-2,Abbott-gw170817}. While the nuclear matter equation of state (EoS) is well known up to nuclear saturation density, characterizing matter at extremely high densities realized in the interior of these stars is a challenge. The nature of the matter at these densities being currently unknown, makes it difficult to create stellar models that would account for the observed findings. Probable candidates include weird quark stars made up of deconfined quarks, hybrid stars made up of deconfined quark cores with hadronic outer shells and pure neutron stars (NSs) made up of hadrons. In the present work, universal relationships for Nss have been investigated. The findings are extremely significant, particularly for the study of dense matter physics \cite{Baym-2018,Alford-2013,Christian-2018,Montana-2019}.

    The first time, a few instances of EoSs that fulfill the maximum mass constraint arisen from the existence of $2~M_{\odot}$ pulsars \cite{Blaschke-2013,Castillo-2013} were presented and were subsequently elaborated by Benic {\it et al.}~\cite{Benic-2015,Castillo-2015}. A systematic Bayesian analysis of the NS EoS with constraints deduced from observations was presented in Ref.~\cite{Castillo-2016}. Some of the recent works dedicated to the study of the NS interior and possible existence of twin stars can be found in Refs.\cite{Ayriyan-2018,Maslov-2019,Alvarez-2017,Bejger-2017,
Bhattacharyya-2010,Christian-2019,Christian-2021,Christian-2022,Han-2019a,
Espinoza-2022,Tan-2022,Li-2021,Sharifi-2021,Benitez-2021,Sendra-2020,
Castillo-2019,Paschalidis-2018,Spinella-2017,Alford-2017,Zacchi-2017,
Zacchi-2016,Bhattacharyya-2005,Sen-2022,Minamikawa-2021,Pietri-2019,
Zdunik-2013,Han-2020,Li-2020,Largani-2022,Ivanytskyi-2022,Contrera-2022,
Schram-2016,Burgio-2018,Sieniawska-2019,Most-2018,Nandi-2018,Deloudis-2021,
Wang-2022,Banik-2004,Banik-2001,Banik-2003,Haensel-2016,Bozzola-2019,
Pereira-2020,Han-2019}.  

    At a frequency of $f=709$ Hz, pulsar PSR J0952-0607 was initially identified by Bassa {\it et al.} \cite{Bassa-2017} as the fastest known spinning pulsar in the disk of Milky Way. Recently, Romani {\it et al.} \cite{Romani-2022} have discovered that the mass of PSR J0952-0607 is $M=2.35\pm 0.17 \ M_{\odot}$ which is the precisely measured largest mass found till date. The aforementioned findings led to investigations into the limits of the evolution of dense nuclear matter and may even reexamine many theoretical predictions about fundamental characteristics of NSs \cite{Ecker-2022} like the presence of twin stars. The twin stars, however, might not exist if there is a supermassive NS \cite{Christian-2021}. A methodical and meticulous examination ought to be contemplated, as the detection of these supermassive stars is crucial in verifying the reliability of the associated theoretical predictions. Conversely, recent discoveries about the radius and tidal deformability of stars with mass close to $1.4~M_{\odot}$ should be taken into account when combining theoretical models for the maximum mass of NSs. A nuclear model is often evaluated based on its ability to accurately estimate the observable radial and tidal deformability values while predicting the maximum achievable mass. Numerous studies have examined the rotating NSs with unusual degrees of freedom in their cores \cite{Bhattacharyya-2005,Banik-2004,Haensel-2016,Bozzola-2019}. 
    
    In the present study, we employ the constraints on mass and radius of the recently observed compact objects, namely, the supernova remnant HESS J1731-347~\cite{Dor22}, the GW170817 event \cite{Abbott19}, PSR J1614-2230 \cite{Arz18}, PSR J0348+0432 \cite{Ant13}, PSR J0740+6620 \cite{Cro20} and PSR J0952-0607 \cite{Rom22} pulsars. The the supernova remnant HESS has $M=0.77_{-0.17}^{+0.20}\ M_{\odot}$ and $R=10.4_{-0.78}^{+0.86} {\rm km}$, while the observations were made with the help of X-ray spectrum modeling and a reliable distance estimate from Gaia observations \cite{Dor22}. The properties of neutron stars, including mass ($M$), radius ($R$), compactness ($C$), Love number ($k_2$), dimensionless tidal deformability ($\Lambda$), dimensionless quadrupole moment($Q$) and the dimensionless moment of inertia ($I$), have been calculated. It has been explored how these quantities are universally related. 

    The paper is arranged in the following manner. In Sec.~\ref{sec:Section 2} we provide basic formalism leading to the tidal deformability while rate of rotation of an inertial frame and quadrupole deformation are described briefly in Sec.~\ref{sec:Section 3}. In Sec.~\ref{sec:Section 4} results of the present calculations are presented and discussed. Finally, summary of the present theoretical endeavor along with concluding remarks are furnished in Sec.~\ref{sec:Section 5}. In the present work we use units $G=c=1$ throughout.  

\begin{figure}[t]
\vspace{-0.0cm}
\eject\centerline{\epsfig{file=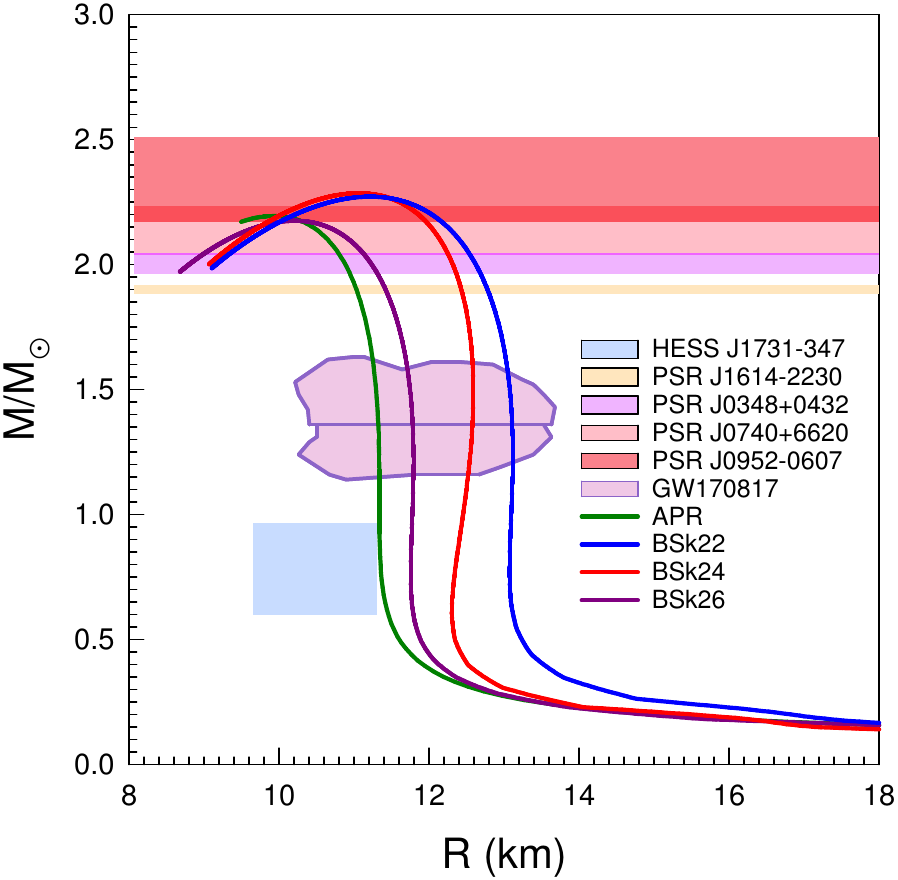,height=8cm,width=8cm}}
\caption{The mass-radius plots for APR, BSk22, BSk24 and BSk26 EoSs. The shaded regions represent the HESS J1731-347 remnant \cite{Dor22}, the GW170817 event \cite{Abbott19}, PSR J1614-2230 \cite{Arz18}, PSR J0348+0432 \cite{Ant13}, PSR J0740+6620 \cite{Cro20}, and PSR J0952-0607 \cite{Rom22} pulsar observations for the possible maximum mass.}
\label{fig1}
\vspace{-0.0cm}
\end{figure}

\begin{figure}[t]
\vspace{0.0cm}
\eject\centerline{\epsfig{file=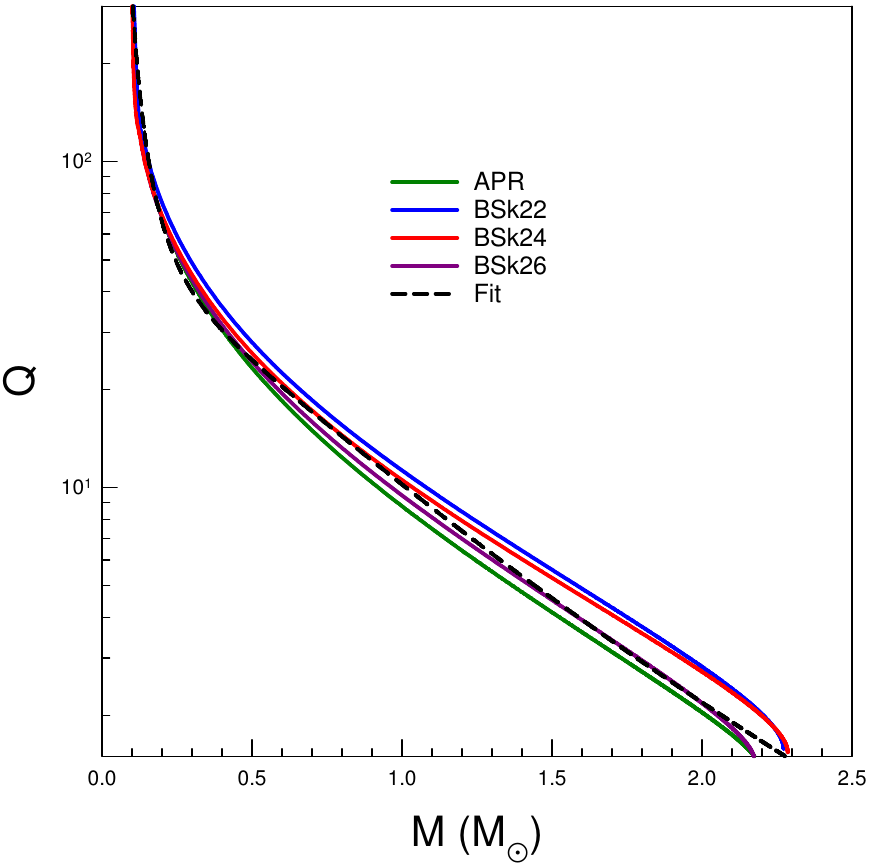,height=8cm,width=8cm}}
\caption{Fitting curve (black dashed), given in Eq.(\ref{L-tild-2}), and numerical results (continuous lines) of the universal relation for dimensionless quadrupole moment ($Q$) vs mass ($M$) for APR, BSk22, BSk24 and BSk26 EoSs.}
\label{fig2}
\vspace{0.0cm}
\end{figure}

\begin{figure}[t]
\vspace{0.0cm}
\eject\centerline{\epsfig{file=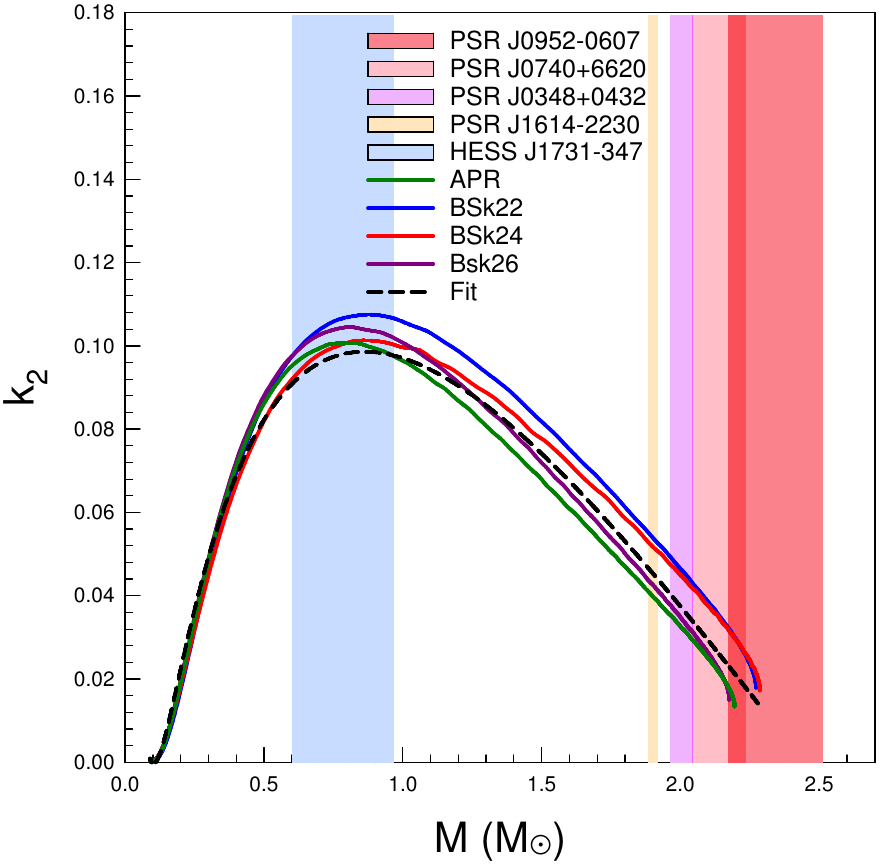,height=8cm,width=8cm}}
\caption{Fitting curve (black dashed), given in Eq.(\ref{L-tild-2}), and numerical results (continuous lines) of the universal relation for Love number ($k_2$) vs mass ($M$) for APR, BSk22, BSk24 and BSk26 EoSs. The shaded regions represent the HESS J1731-347 remnant \cite{Dor22}, the GW170817 event \cite{Abbott19}, PSR J1614-2230 \cite{Arz18}, PSR J0348+0432 \cite{Ant13}, PSR J0740+6620 \cite{Cro20}, and PSR J0952-0607 \cite{Rom22} pulsar observations for the possible maximum mass.}
\label{fig3}
\vspace{-0.14cm}
\end{figure}
\noindent 

\begin{figure}[t]
\vspace{0.0cm}
\eject\centerline{\epsfig{file=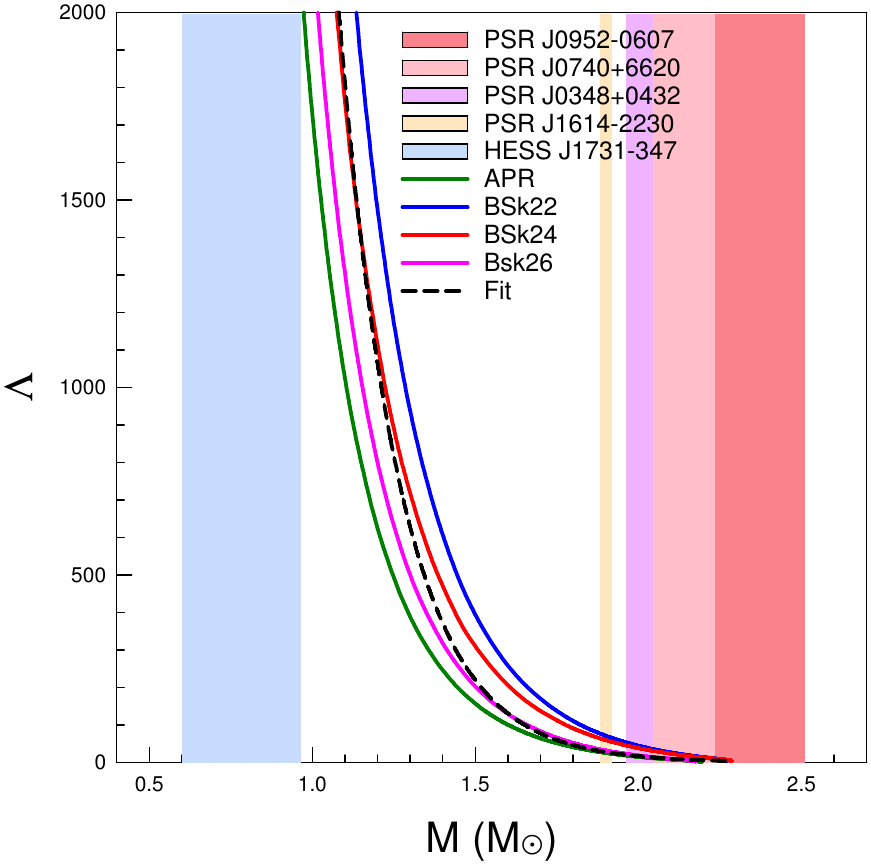,height=8cm,width=8cm}}
\caption
{Fitting curve (black dashed), given in Eq.(\ref{L-tild-2}), and numerical results (continuous lines) of the universal relation for dimensionless tidal deformability ($\Lambda$) vs mass ($M$) for APR, BSk22, BSk24 and BSk26 EoSs. The shaded regions represent the HESS J1731-347 remnant \cite{Dor22}, the GW170817 event \cite{Abbott19}, PSR J1614-2230 \cite{Arz18}, PSR J0348+0432 \cite{Ant13}, PSR J0740+6620 \cite{Cro20}, and PSR J0952-0607 \cite{Rom22} pulsar observations for the possible maximum mass.} 
\label{fig4}
\vspace{-0.14cm}
\end{figure}
\noindent
\noindent
\section{Tidal deformability} 
\label{sec:Section 2}

\subsection{Tolman-Oppenheimer-Volkoff equations} 
\label{subsection 2a}
    The hydrostatic equilibrium of compact star matter is well determined
by a system of differential equations known as the Tolman-Oppenheimer-Volkoff (TOV) equations \cite{TOV39a,TOV39b}. The structure of a slowly rotating, spherical star can, therefore, be well described by the TOV equations  

\beq
\label{tovequation1}
&&\frac{dM(r)}{dr}=4\pi r^2\varepsilon(r), 
\\\nn
&&\frac{d\nu(r)}{dr}=2\frac{[M(r)+4\pi r^3P(r)]}{r[r-2M(r)]}, 
\\\nn
\label{tovequation3}
&&\frac{dP(r)}{dr}=-\frac{[P(r)+\varepsilon(r)]}{2} \frac{d\nu(r)}{dr}.
\eeq
where $\varepsilon(r)$ and $P(r)$ are the energy density and pressure distribution inside the compact star. Since all functions considered here depend on radial co-ordinate $r$, this dependency will be omitted henceforth. The quantities, $R$ and $M(R)$, hereafter will indicate, respectively, the radius and the mass of the compact star, obtained by solving Eqs.(\ref{tovequation1}) for a designated EoS with the boundary condition that $P(r) \rightarrow 0$ on the surface of the compact star at $r=R$.

\begin{figure}[t]
\vspace{0.0cm}
\eject\centerline{\epsfig{file=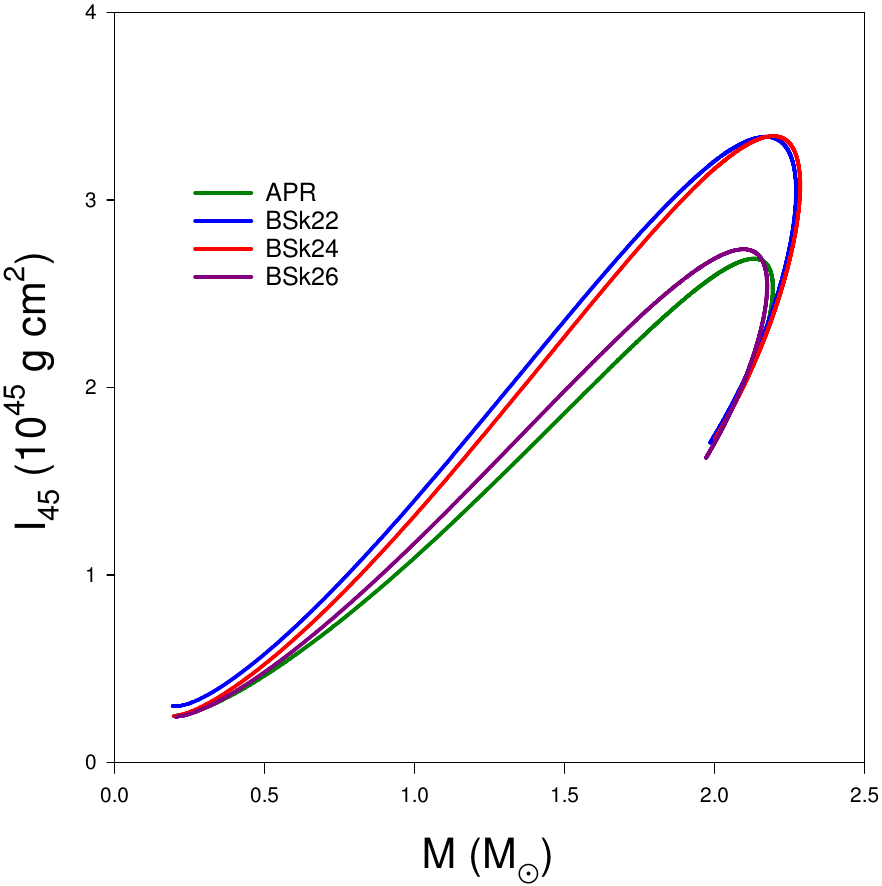,height=8cm,width=8cm}}
\caption{Plots of moment of inertia $I_{45}$ in units of 10$^{45}$ g cm$^2$ vs $M$ for APR, BSk22, BSk24 and BSk26 EoSs.} 
\label{fig5}
\vspace{0.0cm}
\end{figure}

\begin{figure}[htbp]
\vspace{0.0cm}
\eject\centerline{\epsfig{file=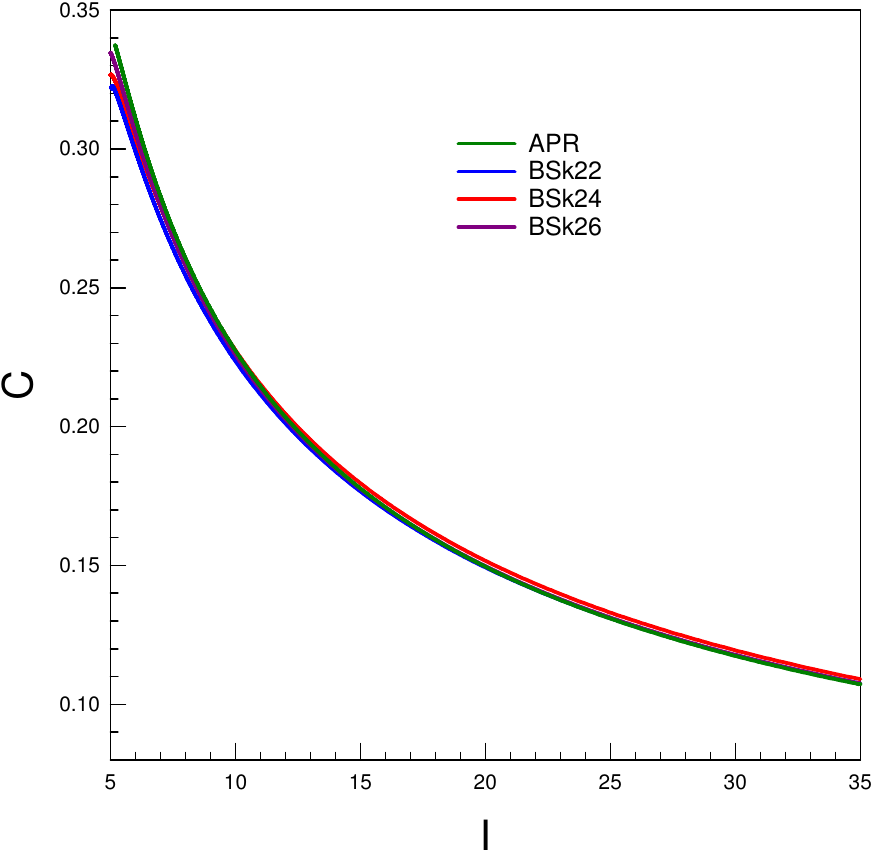,height=8cm,width=8cm}}
\caption{Plots of compactness ($C$) vs dimensionless moment of inertia ($I$) for APR, BSk22, BSk24 and BSk26 EoSs.} 
\label{fig6}
\vspace{0.0cm}
\end{figure}
    
\begin{figure}[htbp]
\vspace{0.0cm}
\eject\centerline{\epsfig{file=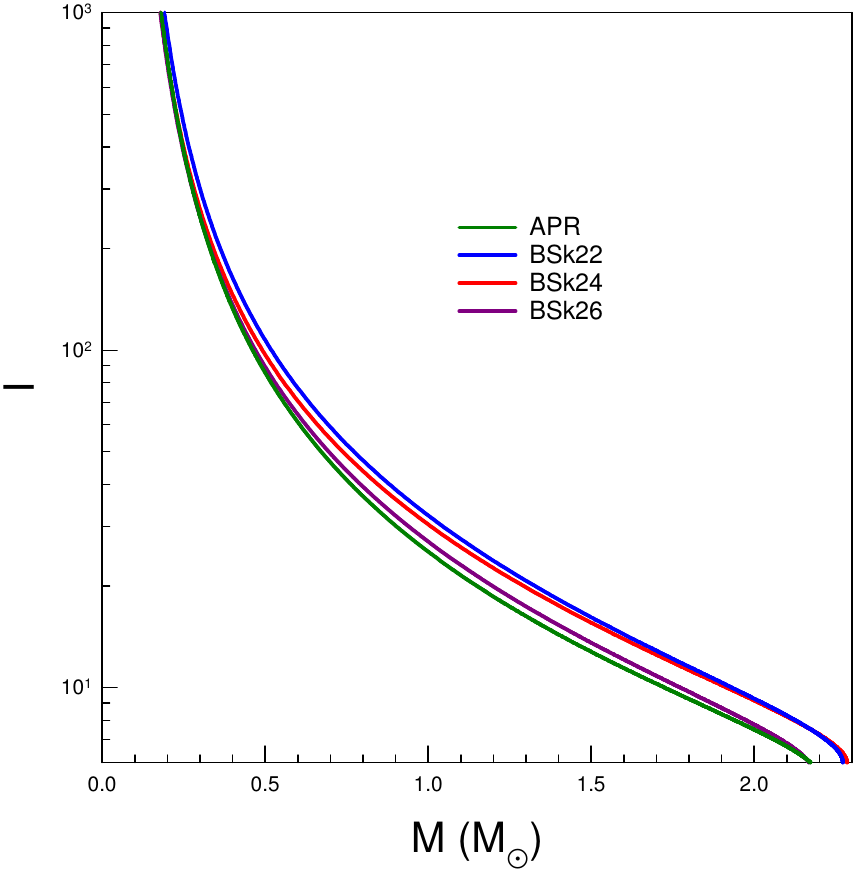,height=8cm,width=8cm}}
\caption{Plots of dimensionless moment of inertia ($I$) vs mass ($M$) for APR, BSk22, BSk24 and BSk26 EoSs.} 
\label{fig7}
\vspace{0.0cm}
\end{figure}

\subsection{Love number and tidal deformability} 
\label{subsection 2b}

    The GW emission during the late phase of inspiraling of a binary NS system ahead of merger \cite{Pos2010,Fla2008,Hin2008} is an important source for the GW detectors. In the inspiral phase the tidal effects can be detected \cite{Fla2008}. In recent years, observations of GWs resulting from sources like the merger of black hole–NS and NS-NS binary systems induce measurements of various properties of NSs.

    The quadrupole moment $Q_{ij}$ of compact star and the external tidal field $\mathcal{E}_{ij}$ are defined as coefficients in an asymptotic expansion of the total metric at large distances $r$ from the star. The tidal Love number $k_2$ depends upon EoS and characterizes the response of NS to the tidal field  $\mathcal{E}_{ij}$ \cite{Fla2008}. This relation can be expressed as 
    
\begin{equation}
Q_{ij} = -\frac{2}{3}k_2R^5\mathcal{E}_{ij}= -\lambda\mathcal{E}_{ij}
\label{eq:lambda}
\end{equation}
where $R$ is radius of NS and $\lambda=2k_2R^5/3$ is the tidal deformability. The $l=2$ tidal Love number is given by \cite{Fla2008,Hin2008}

\begin{eqnarray}
k_2 &=& \frac{8C^5}{5}(1-2C)^2[2+2C(y-1)-y]\nonumber\\
      & & \times\bigg\{2C[4(1+y)C^4+(6y-4)C^3+(26-22y)C^2\nonumber\\
      & & +3(5y-8)C-3y+6]+3(1-2C)^2\nonumber\\
      & & \times[2C(y-1)-y+2] \ln(1-2C)\bigg\}^{-1},\nonumber\\
\label{eq:k2}
\end{eqnarray}     
where the dimensionless quantity $C=M(R)/R$ is the compactness of the star and the dimensionless quantity $y$ is defined as
 
\begin{equation}
y = \frac{ R\, \beta(R)} {H(R)}
\end{equation}
for the internal solution determined by numerically solving the second-order differential equation for $H(r)$. The second-order differential equation for $H$ can be separated into a first-order system of ordinary differential equations in terms of the usual TOV quantities $M$, $p(r)$, and $\varepsilon(p)$, as well as the additional functions $H(r)$, $\beta(r) =
dH/dr$, and the EoS function $f(p) = d\varepsilon/dp$:

\beq \frac{dH}{dr}&=& \beta\\
\frac{d\beta}{dr}&=&2 \left(1 - 2\frac{M}{r}\right)^{-1} H\left\{-2\pi
  \left[5\varepsilon+9
    p+f(\varepsilon+p)\right]\phantom{\frac{3}{r^2}} \right. \nonumber\\
&& \quad \left. +\frac{3}{r^2}+2\left(1 - 2\frac{M}{r}\right)^{-1}
  \left(\frac{M}{r^2}+4\pi r p\right)^2\right\}\nonumber\\
&&+\frac{2\beta}{r}\left(1 -
  2\frac{M}{r}\right)^{-1}\left\{-1+\frac{M}{r}+2\pi r^2
  (\varepsilon-p)\right\}.\nonumber\\\eeq
The above equations combined with TOV Eqs.(\ref{tovequation1}) is then solved simultaneously. The system is integrated outward starting just outside the center using the expansions $H(r)=a_0 r^2$ and $\beta(r)=2a_0r$ as $r \to 0$.  The constant $a_0$ determines how much the star is deformed and can be chosen
arbitrarily as it cancels in the expression for the Love number. The boundary conditions that determine the unique choice of this solution follow from matching the interior and exterior solutions and their first derivatives at
the boundary of the star, where $r=R$. The dimensionless tidal deformability related to the Love number can now be given by 

\begin{equation}
\Lambda=\frac{2}{3}k_2 C^{-5}.
\label{Lamb-1}
\end{equation}
where $C$ is the compactness of the star defined previously. One of the primary quantities that can be accurately determined by the detection of the associated GWs is the tidal deformability, or $\Lambda$.

\section{Rate Rotation of Inertial Frame}
\label{sec:Section 3}

    The compact star has been assumed to be a slowly rotating object having a uniform angular velocity $\Omega$ \cite{Hartle67,Ar77}. To the first order in $\Omega$ the shape of the star remains spherical and there is only one function, $\omega$, to be determined, which is the angular velocity of the local inertial frame \cite{Hartle67,Hartle68,Hartle73}. By introducing the function $\varpi=\Omega-\omega$, the angular velocity of a point in the star measured with respect to the angular velocity of the local inertial frame, it can be shown that it  satisfies the following equation \cite{Hartle68,Benhar2005}
    
\begin{figure}[ht!]
\vspace{0.0cm}
\eject\centerline{\epsfig{file=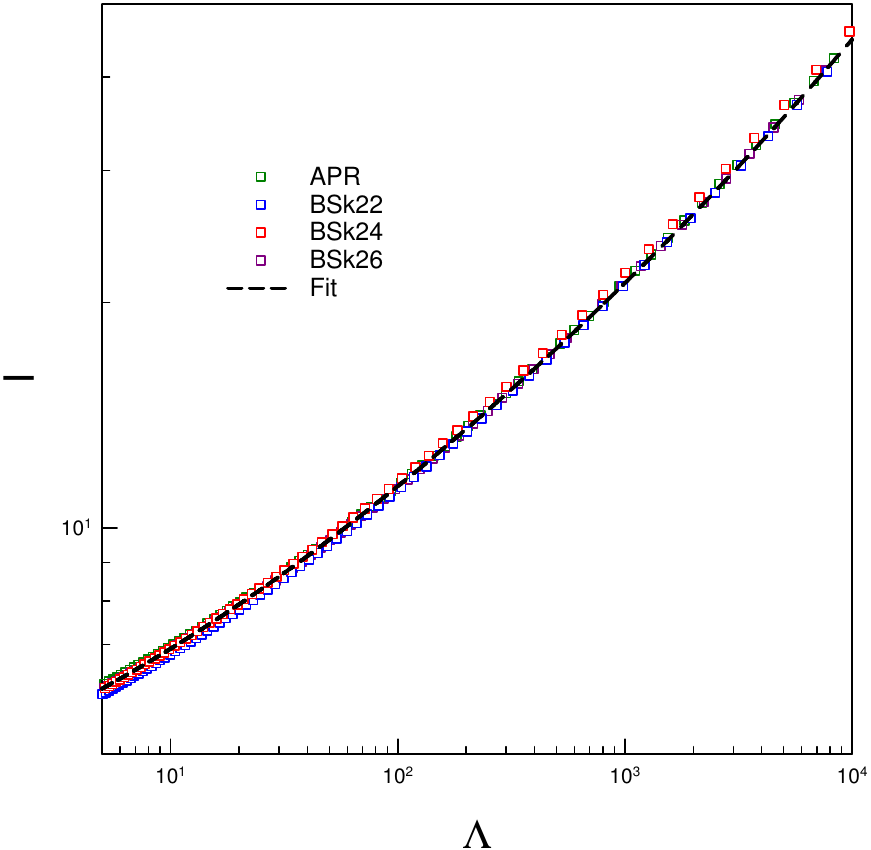,height=8cm,width=8cm}}
\caption{Fitting curve (black dashed), given in Eq.(\ref{L-tild-2}), and numerical results (points) of the universal relation for dimensionless moment of inertia ($I$) vs dimensionless tidal deformability ($\Lambda$) for APR, BSk22, BSk24 and BSk26 EoSs.} 
\label{fig8}
\vspace{0.0cm}
\end{figure}
    
\begin{figure}[t]
\vspace{0.0cm}
\eject\centerline{\epsfig{file=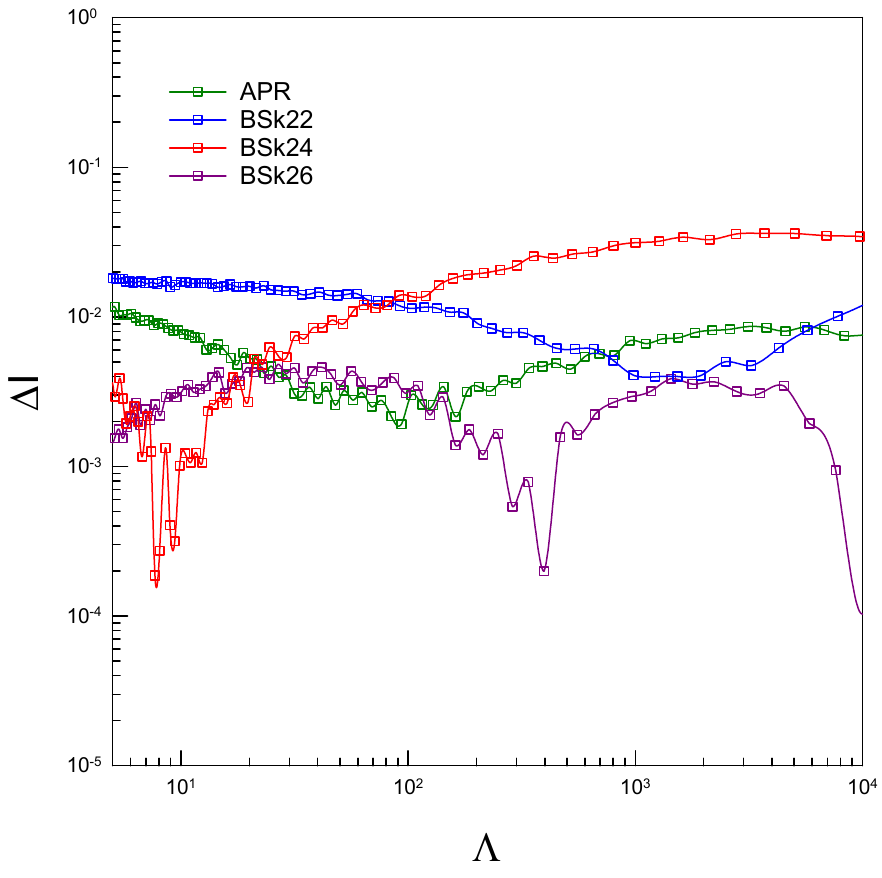,height=8cm,width=8cm}}
\caption{Relative fractional errors between the fitting curve and numerical results of Fig.-8.} 
\label{fig9}
\vspace{0.2cm}
\end{figure}

\begin{figure}[t]
\vspace{0.0cm}
\eject\centerline{\epsfig{file=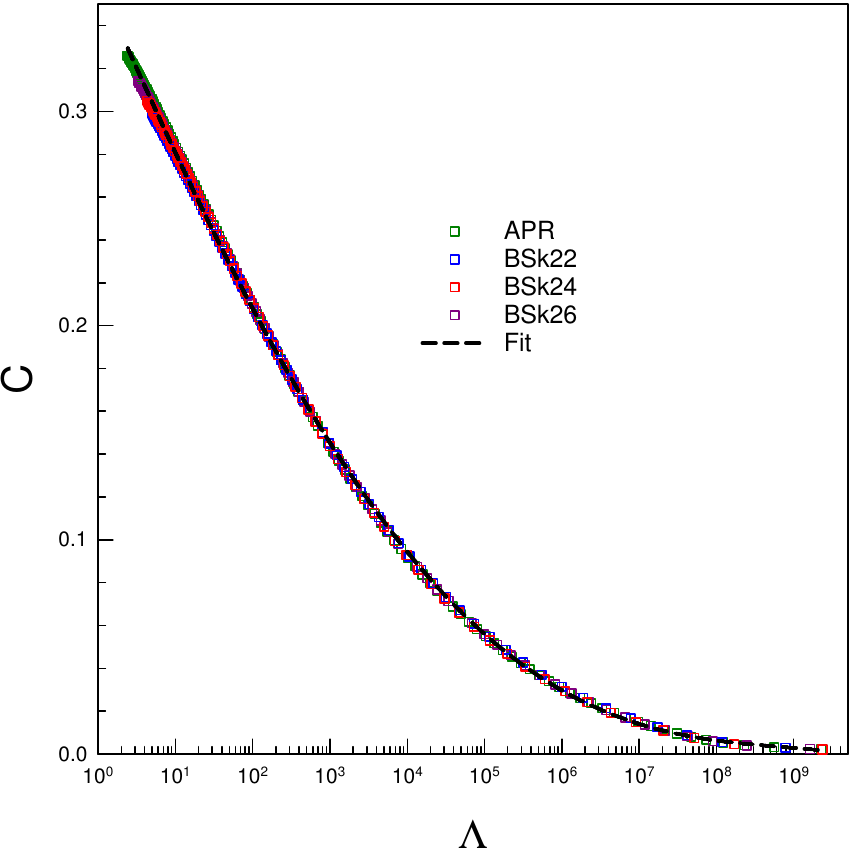,height=8cm,width=8cm}}
\caption{Fitting curve (black dashed), given in Eq.(\ref{L-tild-2}), and numerical results (points) of the universal relation for compactness ($C$) vs dimensionless tidal deformability ($\Lambda$) for APR, BSk22, BSk24 and BSk26 EoSs.} 
\label{fig10}
\vspace{-0.07cm}
\end{figure}

\begin{figure}[t]
\vspace{0.0cm}
\eject\centerline{\epsfig{file=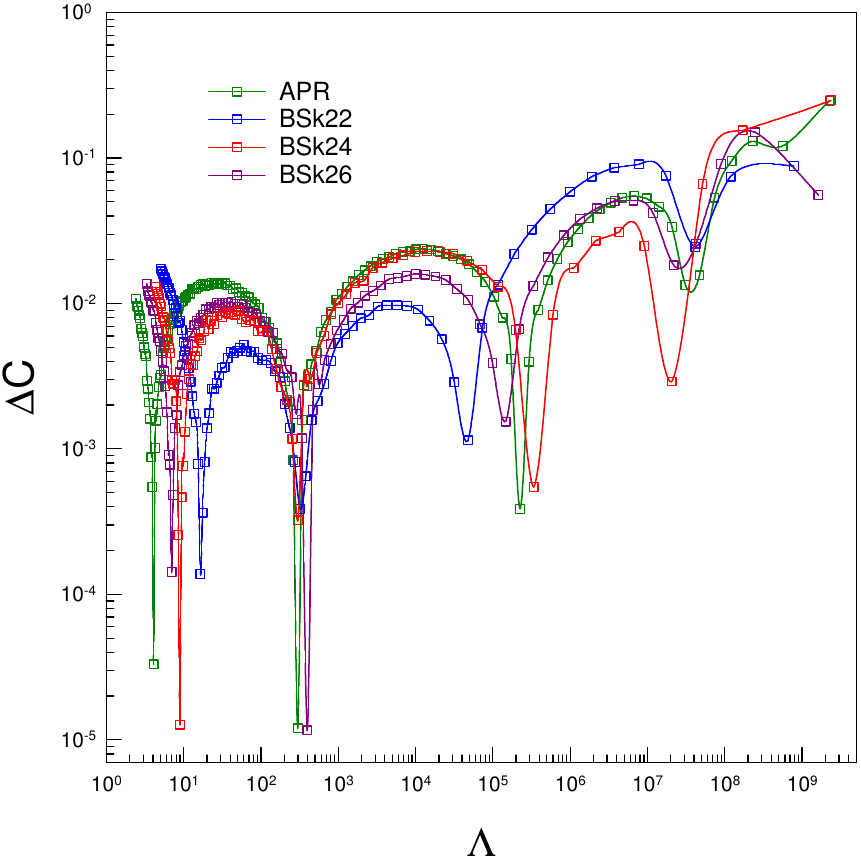,height=8cm,width=8cm}}
\caption{Relative fractional errors between the fitting curve and numerical results of Fig.-10.} 
\label{fig11}
\vspace{0.0cm}
\end{figure}
    
\be
\frac{1}{r^4}\frac{d}{dr}(r^4j\frac{d\varpi}{dr})+\frac{4}{r}
\frac{dj}{dr}{\varpi}=0,
\label{eqpai}
\ee
where $\label{defj} j(r)= e^{-\nu/2}\sqrt{1-\frac{2 M}{r}}$ which $\to 1$ at $r \to R$. The Eq.(\ref{eqpai}) can be solved subject to the boundary conditions that $\varpi(r)$ is regular as $r \to 0$ and $\varpi(r) \to \Omega$ as $r \to \infty$. The moment of inertia of the star can then be calculated dividing the total angular momentum $J$ by $\Omega$ as 

\vspace{-0.0cm}
\begin{equation}
\mathcal{I} = \frac{1}{6\Omega} R^4 \frac{d\varpi}{dr}\Big|_{r= R}.
\label{seqn10}
\end{equation}
\noindent
In the dimensionless form moment of inertia $I$ of the star is given by $I=\mathcal{I}/M^3$.

\begin{figure}[t]
\vspace{0.0cm}
\eject\centerline{\epsfig{file=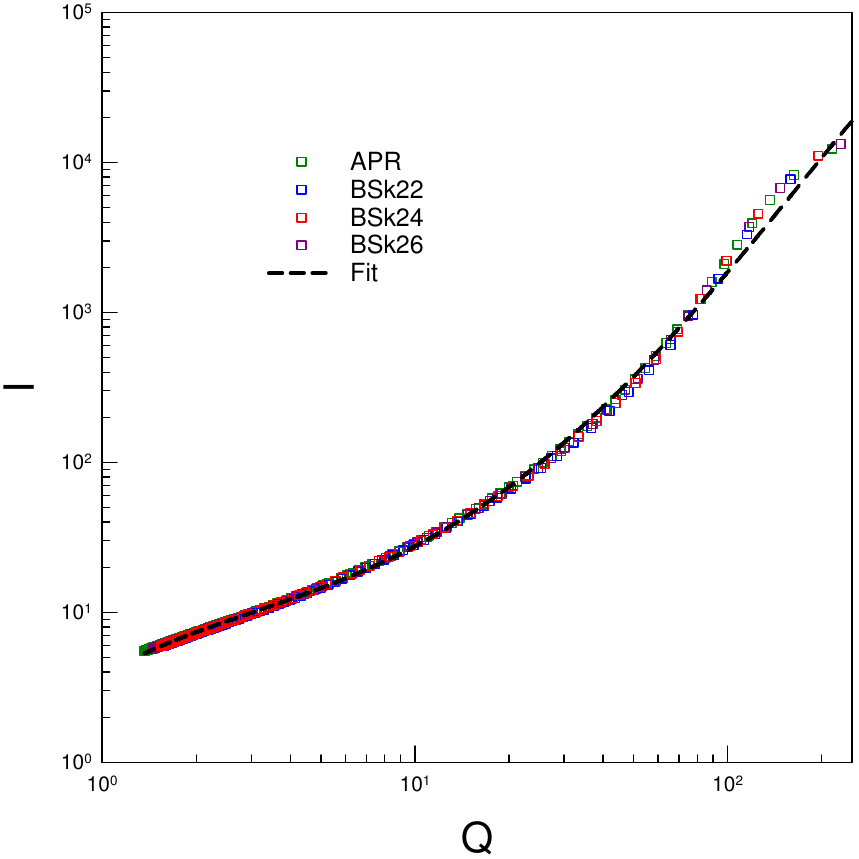,height=8cm,width=8cm}}
\caption{Fitting curve (black dashed), given in Eq.(\ref{L-tild-2}), and numerical results (points) of the universal relation for dimensionless moment of inertia ($I$) vs dimensionless quadrupole moment ($Q$) for APR, BSk22, BSk24 and BSk26 EoSs.} 
\label{fig12}
\vspace{-0.07cm}
\end{figure}

\begin{figure}[t]
\vspace{0.0cm}
\eject\centerline{\epsfig{file=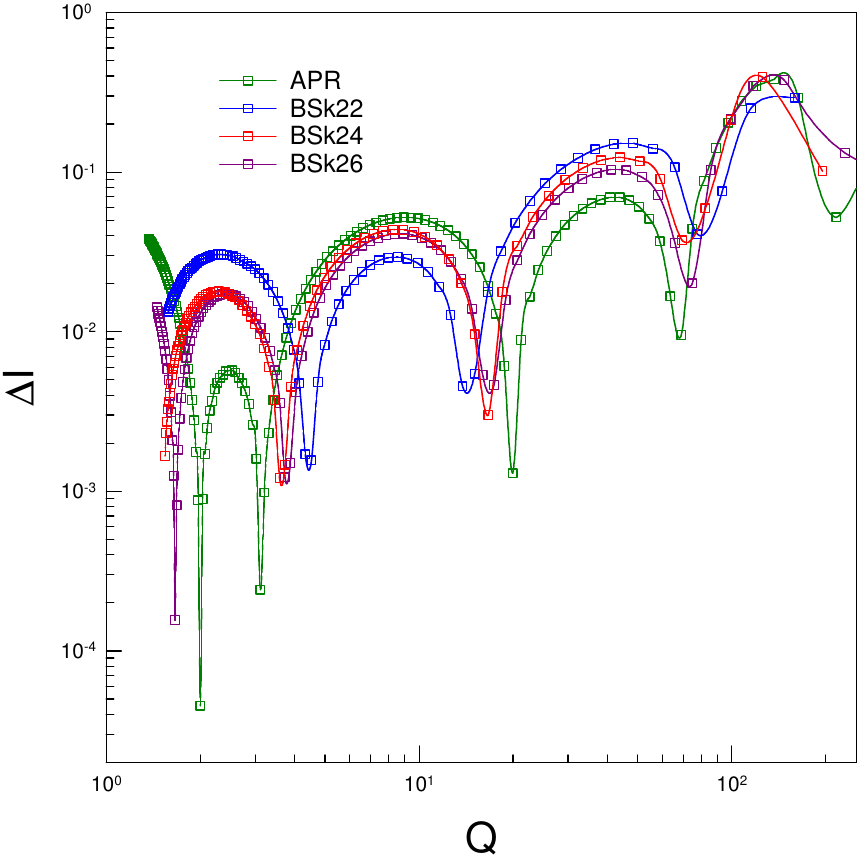,height=8cm,width=8cm}}
\caption{Relative fractional errors between the fitting curve and numerical results of Fig.-12.} 
\label{fig13}
\vspace{-0.07cm}
\end{figure}		
 
\begin{figure}[t]
\vspace{0.0cm}
\eject\centerline{\epsfig{file=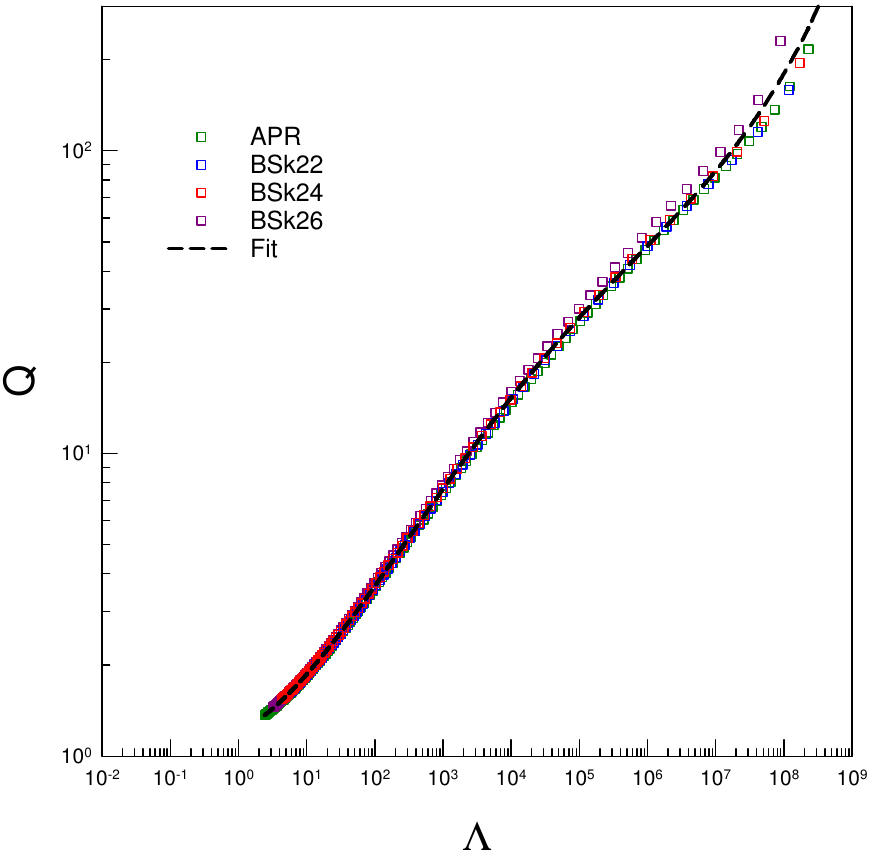,height=8cm,width=8cm}}
\caption{Fitting curve (black dashed), given in Eq.(\ref{L-tild-2}), and numerical results (points) of the universal relation for dimensionless quadrupole moment ($Q$) vs dimensionless tidal deformability ($\Lambda$) for APR, BSk22, BSk24 and BSk26 EoSs.} 
\label{fig14}
\vspace{-0.07cm}
\end{figure}

\begin{figure}[t]
\vspace{0.0cm}
\eject\centerline{\epsfig{file=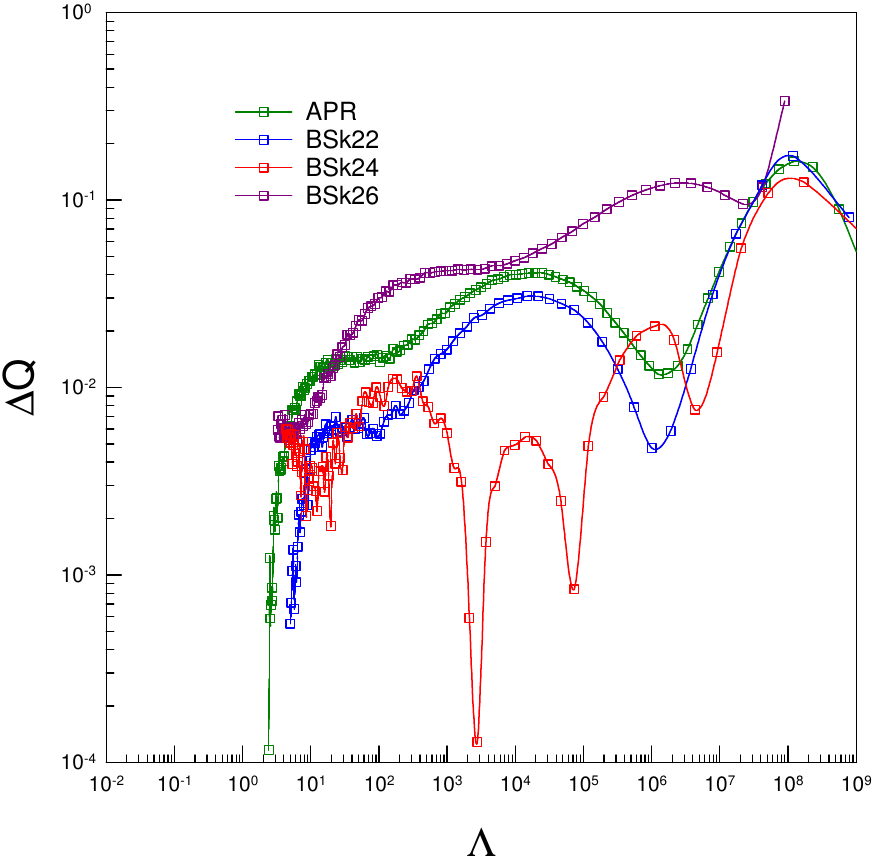,height=8cm,width=8cm}}
\caption{Relative fractional errors between the fitting curve and numerical results of Fig.-14.} 
\label{fig15}
\vspace{-0.07cm}
\end{figure}

    The Eq.(\ref{eqpai}) can be reduced to two first order equations by defining two auxiliary functions of $r$

\be
\label{defuchi}
\chi=j\varpi,\qquad\qquad
u=r^4j\frac{d\varpi}{dr}
\ee
that satisfy
\beq
r \le R\quad&&\frac{d\chi}{dr}=\frac{u}{r^4}-\frac{4\pi 
\nonumber
r^2(\varepsilon+P)\chi}{r-2M}\\
&&\frac{du}{dr}=\frac{16\pi r^5(\varepsilon+P)\chi}{r-2M} .
\label{eqpai1}
\eeq
The above equations have to be integrated outward from the center of the star $r=0$ to its surface at $r=R$ imposing ensuing boundary (initial) conditions that

\be
\chi(0)=j(0)\varpi_c,\qquad\quad
u(0)=0.
\label{cond1}
\ee
These initial conditions result from the behavior of $\varpi$ near $r \to 0$
 
\be
\label{varp}
\varpi \sim \varpi_c(1 + \varpi_2 r^2 +....),
\ee
where $\varpi_c$ is a constant. The constant $\varpi_c$ can be chosen
arbitrarily. When one reaches the surface and only then one can determine the angular velocity $\Omega$ and angular momentum $J$, corresponding to $\varpi_c$. It is worthwhile to mention here that from Eq.(\ref{defuchi}) and Eq.(\ref{varp}) it follows that the entity $u$ goes to zero more rapidly than $r^4$.

    For $r \ge R$, $P=0$, $\varepsilon=0$, $M \equiv M(R)$, $j(R)=1$ and
the solution of Eqs.(\ref{eqpai1}) can be given by

\be
\label{uchiout}
r \ge R\qquad \chi(r)=\Omega-\frac{2J}{r^3},\qquad
u(r)=6J,
\ee
where $J$ is a constant of motion that, to first order in $\Omega$, reflects the star's angular momentum. The requirement of continuity of the exterior and interior solutions demands that they be matched at $R$, which yields the values of $\Omega$ and $J$. This indicates that $J=u(R)/6$ and $\Omega=\chi(R)+2J/R^3$. If a different value of $\Omega$ is desired, one rescales the function $\varpi(r)$ through $\varpi_c$ to obtain it.

\subsection{Quadrupole deformation}
\label{Subsection 3a}
    The equations for $h_2(r)$ and $v_2(r)$ are given by

\beq
\label{quad}
\frac{dv_2}{dr}&=&-\frac{d\nu}{dr}h_2+(\frac{1}{r}+
\frac{1}{2}\frac{d\nu}{dr})\left[
\frac{8\pi r^5(\varepsilon+P)\chi^2}{3(r-2M)}+\frac{u^2}{6r^4}\right]
\\\nonumber
\frac{dh_2}{dr}&=&\left[-\frac{d\nu}{dr}+
\frac{r}{r-2M}(\frac{d\nu}{dr})^{-1}(8\pi(\varepsilon+P)-
\frac{4M}{r^3})\right]h_2\\
\nonumber
&-&\frac{4v_2}{r(r-2M)}(\frac{d\nu}{dr})^{-1}+
\frac{u^2}{6r^5}\left[
\frac{1}{2}\frac{d\nu}{dr}r-\frac{1}{r-2M}(\frac{d\nu}{dr})^{-1}\right]
\\\nonumber
&+&\frac{8\pi r^5(\varepsilon+P)\chi^2}{3r(r-2M)}\left[
\frac{1}{2}\frac{d\nu}{dr}r+
\frac{1}{r-2M}(\frac{d\nu}{dr})^{-1}\right].
\eeq
which are responsible for the quadrupole ($l=2$) deformation of a compact star. In order to solve these equations for $r \le R$, it is important to note that a regular solution of Eqs.(\ref{quad}) at $r \to 0$ must behave as

\begin{figure}[t]
\vspace{0.0cm}
\eject\centerline{\epsfig{file=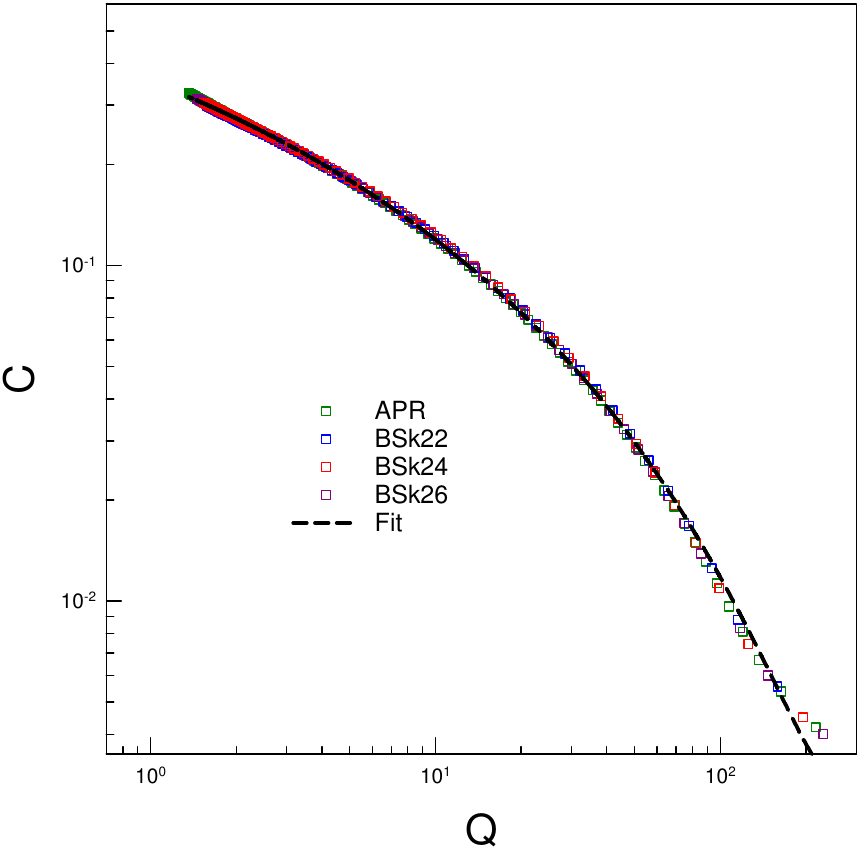,height=8cm,width=8cm}}
\caption{Fitting curve (black dashed), given in Eq.(\ref{L-tild-2}), and numerical results (points) of the universal relation for compactness ($C$) vs dimensionless quadrupole moment ($Q$) for APR, BSk22, BSk24 and BSk26 EoSs.} 
\label{fig16}
\vspace{-0.07cm}
\end{figure}

\begin{figure}[t]
\vspace{0.0cm}
\eject\centerline{\epsfig{file=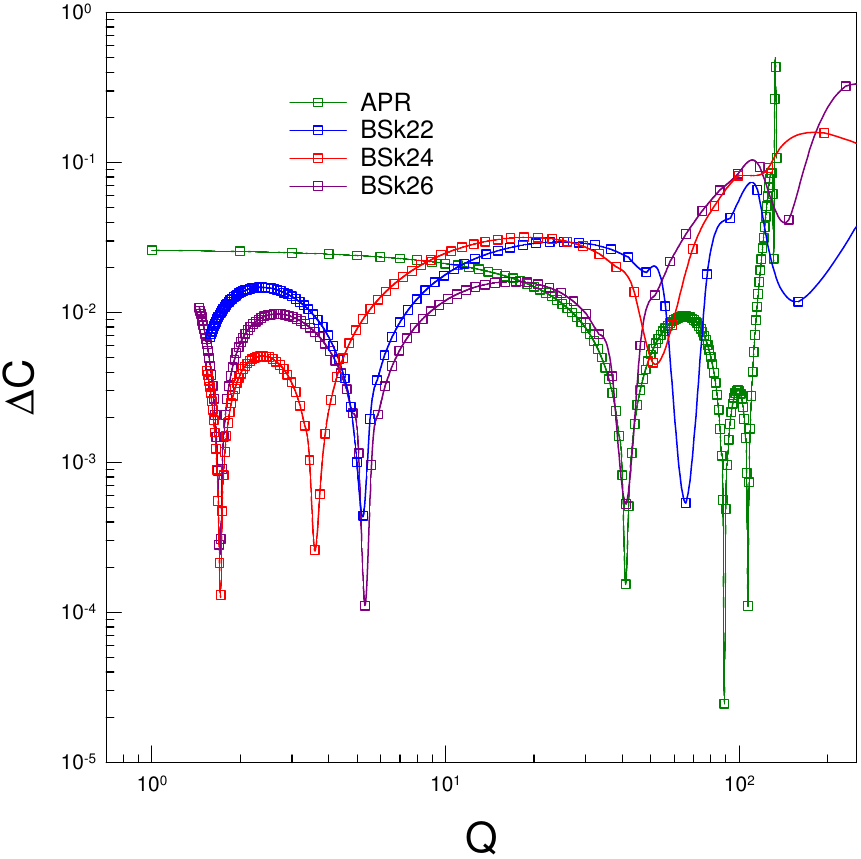,height=8cm,width=8cm}}
\caption{Relative fractional errors between the fitting curve and numerical results of Fig.-16.} 
\label{fig17}
\vspace{-0.07cm}
\end{figure}     

\be
r \to 0\qquad h_2(r) \sim A r^2,\qquad v_2 \sim B r^4,
\label{asint2}
\ee
where the equation that describes how the constants $A$ and $B$ are connected is given by

\be
\label{asint3}
B + 2\pi \left[\frac{1}{3}\varepsilon(0)+P(0) \right] A 
= \frac{2}{3}\pi \left[\varepsilon(0)+P(0) \right] \left(j(0)\varpi_c\right)^2.
\ee
Following Refs.\cite{Hartle68,Benhar2005}, the general solution of Eqs.(\ref{quad}) for $r \le R$ can be written as

\beq
\label{gen1}
&&h_{2}(r)=h^P_2 + C_2~h^H_{2}\\\nonumber
&&v_{2}(r)=v^P_2 + C_2~v^H_{2}.
\eeq 
The constant $C_2$ remains to be determined while $(h_2^P,v_2^P)$ are a particular solution which can be found by integrating Eqs.(\ref{quad}) with the initial  conditions, say,  $A=1$ and $B$ given  by Eq.(\ref{asint3}) and $(h_2^H,v_2^H)$ are the solution  of the homogeneous equations which can be obtained by setting $\chi=0$ and $u=0$ in Eqs.(\ref{quad}) and integrating numerically with the following boundary conditions

\beq
r \to 0
&&h^H_{2}(r)\sim r^2\\\nonumber    
&&v^H_{2}(r)\sim -2\pi~ \left[\frac{1}{3}\varepsilon(0)+P(0) \right] r^4.\nonumber
\eeq

\begin{figure}[t]
\vspace{0.0cm}
\eject\centerline{\epsfig{file=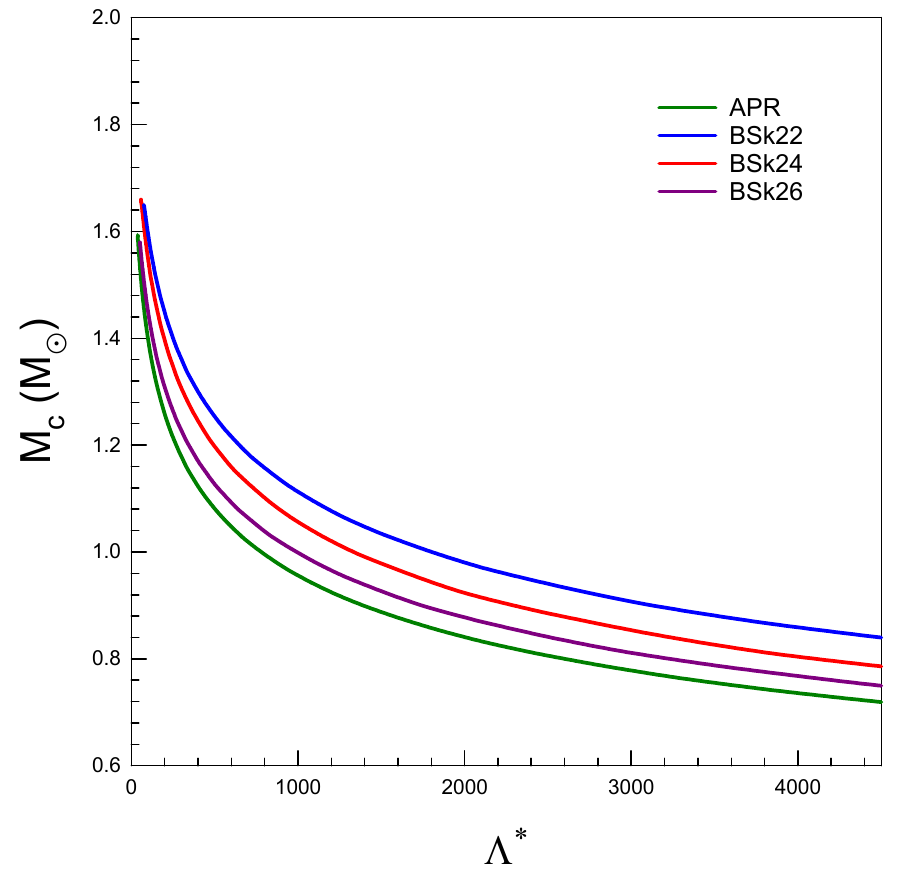,height=8cm,width=8cm}}
\caption{Plots of chirp mass versus effective tidal deformability ($\Lambda^*$) for binary mass
ratio $q=0.7$.} 
\label{fig18}
\vspace{-0.07cm}
\end{figure}

\begin{figure}[t]
\vspace{0.0cm}
\eject\centerline{\epsfig{file=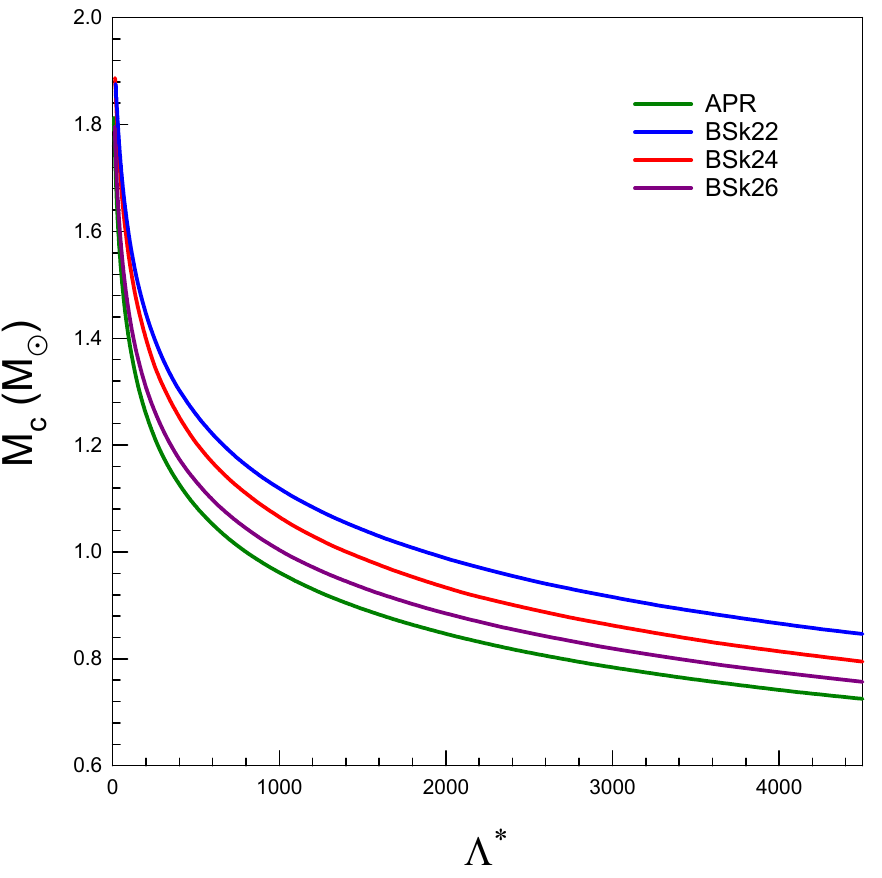,height=8cm,width=8cm}}
\caption{Plots of chirp mass versus effective tidal deformability ($\Lambda^*$) for binary mass
ratio $q=0.9$.} 
\label{fig19}
\vspace{-0.07cm}
\end{figure}

\begin{figure}[t]
\vspace{0.0cm}
\eject\centerline{\epsfig{file=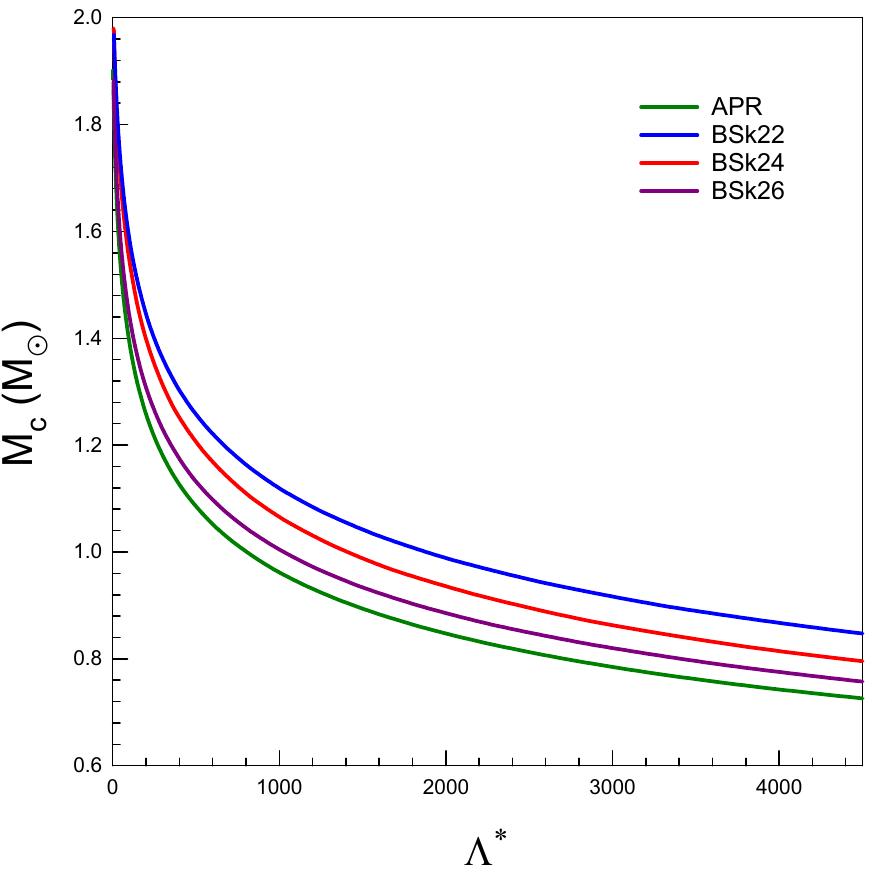,height=8cm,width=8cm}}
\caption{Plots of chirp mass versus effective tidal deformability ($\Lambda^*$) for binary mass
ratio $q=0.99$.} 
\label{fig20}
\vspace{-0.07cm}
\end{figure}

\begin{figure}[t]
\vspace{0.0cm}
\eject\centerline{\epsfig{file=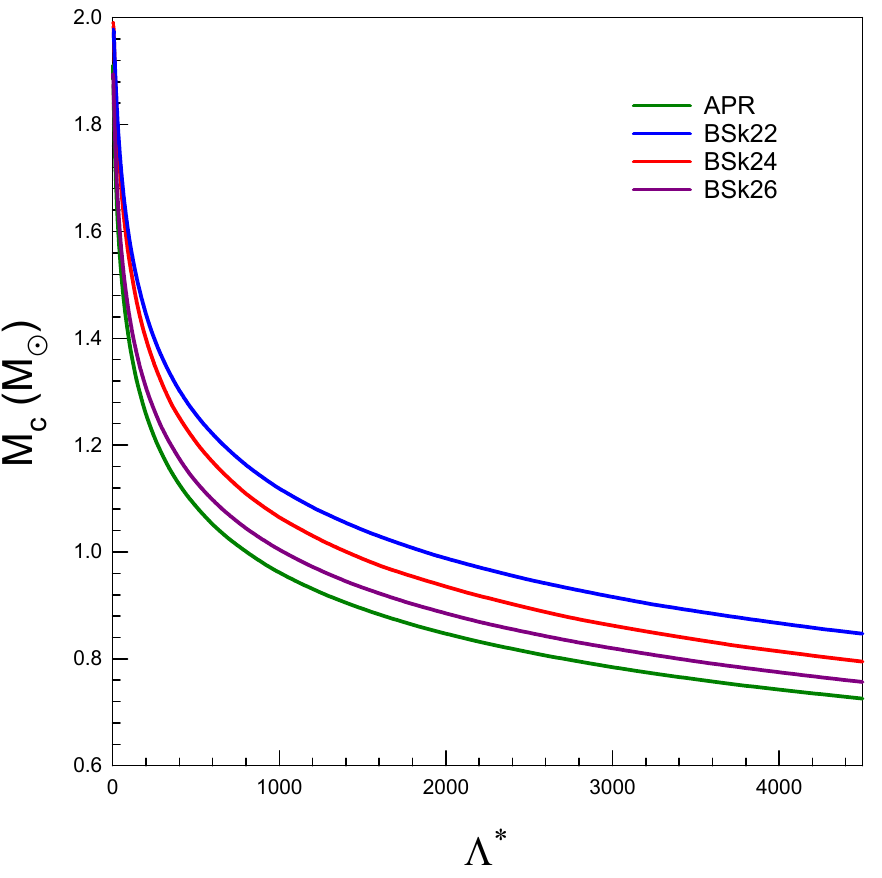,height=8cm,width=8cm}}
\caption{Plots of chirp mass versus effective tidal deformability ($\Lambda^*$) for binary mass
ratio $q=1.0$.} 
\label{fig21}
\vspace{-0.07cm}
\end{figure}

    For $r \ge R$, analytic solutions for Eqs.(\ref{quad}) in terms of the associated Legendre functions of second kind can be deduced \cite{Hartle67,Hartle68} which are

\beq
\label{extgen2}
r \ge R
&&h_2(r)=J^2(\frac{1}{M(R) r^3}+\frac{1}{r^4})+K~Q^2_2(\xi)\\\nonumber     
&&v_2(r)=-\frac{J^2}{r^4}+ K~ \frac{2M(R)}{[r(r-2M(R))]^{1/2}}Q^1_2(\xi)
\eeq
where $K$ is a constant and
\beq
\label{defq}
&&Q^2_2(\xi)=[\frac{3}{2}(\xi^2-1)\log(\frac{\xi+1}{\xi-1})-
\frac{3\xi^3-5\xi}{\xi^2-1}]\\\nonumber
&&Q^2_1(\xi)=(\xi^2-1)^{1/2}[\frac{3\xi^2-2}{\xi^2-1}-
\frac{3}{2}\xi \log(\frac{\xi+1}{\xi-1})].
\eeq
with
\be
\label{defxi}
\xi=\frac{r}{M(R)} -1. 
\ee

    The constants $C_2$ and $K$ can now be evaluated by imposing the condition that the solutions (\ref{gen1}) and the solutions (\ref{extgen2}) are continuous at $r=R$ and hence equal. Therefore, the functions  $h_2$ and $v_2$ are, thus, completely determined.

\noindent

\noindent
\subsection{Quadrupole moment} 
\label{Subsection 3b}

     Once the constants $C_2$ and $K$ have been determined by imposing the continuity condition that the solutions (\ref{gen1}) and the solutions (\ref{extgen2}) be matched and equated at $r=R$, the quadrupole moment $\mathcal{Q}$ can be determined in terms of $K$ as

\be
\mathcal{Q}=\frac{J^2}{M}+\frac{8}{5}K M^3
\label{QDL1}
\ee     
     
which can be expressed in the dimensionless form as

\be
Q=\frac{\mathcal{Q}}{M^3(J/M^2)^2}=1+\frac{8}{5}K\frac{M^4}{J^2}
\label{QDL2}
\ee
\section{Theoretical Calculations and Results} 
\label{sec:Section 4}

    In this work, EoSs obtained from the effective nucleon-nucleon (NN) interaction described by APR \cite{Ak98} and Brussels-Montreal Skyrme effective interaction with BSk22, BSk24 and BSk26 parameter sets \cite{Gor13} have been used. The $\beta$-equilibrated $npe\mu$ neutron star matter has been used for subsequent calculations. The calculations for masses and radii have been performed using FMT \cite{FMT}, BPS \cite{Ba71} and BBP \cite{Ba71a} the EoSs up to the number density of 0.0582 fm$^{-3}$ for the outer crust and the $\beta$-equilibrated NS matter beyond covering the inner crustal and core regions of a compact star. The location of the inner edge of the NS crust, the core-crust transition density and pressure were determined \cite{La24} by dynamical method. The observations of the binary millisecond pulsar J1614-2230 suggest that its mass lies in the range $1.97\pm0.04$ M$_\odot$ \cite{Arz18,De10}. The measurements of radio timing for pulsar PSR J0348+0432 and its companion (white dwarf) have confirmed the mass of the pulsar to be in the range of 1.97$-$2.18 $M_\odot$ \cite{Ant13}. Very recently, the studies for PSR J0740+6620 \cite{Fon21} and for PSR J0952-0607 \cite{Rom22} find masses of 2.08 $\pm$ 0.07 $M_\odot$ and 2.35 $\pm$ 0.17 $M_\odot$, respectively. Current observations of PSR J0740+6620 also suggest its mass to be 2.072$^{+0.067}_{-0.066}$ $M_\odot$ \cite{Leg21,Ril21}. The mass versus radius of NSs obtained using APR, BSk22, BSk24 and BSk26 EoSs have been plotted in Fig.-\ref{fig1}. The maximum masses obtained using APR, BSk22, BSk24 and BSk26 are, repectively, 2.19 $M_\odot$, 2.27 $M_\odot$, 2.28 $M_\odot$ and 2.18 $M_\odot$ \cite{La24} which provide reasonably good estimates for latest observations of NS masses. The shaded regions represent the HESS J1731-347 remnant \cite{Dor22}, the GW170817 event \cite{Abbott19}, PSR J1614-2230 \cite{Arz18}, PSR J0348+0432 \cite{Ant13}, PSR J0740+6620 \cite{Cro20}, and PSR J0952-0607 \cite{Rom22} pulsar observations for the possible maximum mass. 		  
\noindent
\subsection{Universal relationships} 
\label{Subsection 4a}

    The universal relations among $M$, $I$, $\Lambda$, $C$ and $Q$ have been investigated for APR, BSk22, BSk24 and BSk26 EoSs. It has been observed that these relations hold universally for NS sequences, essentially independently of their EoS. Such relations can be numerically fitted with a polynomial given by
    
\begin{equation}
Y_{fit} = \sum_{n=0}^4 a_n X^n 
\label{L-tild-2}
\end{equation}
\noindent 
where the variables and fitted coefficients have been summarized in Table-I. The relative fractional error $\Delta Y$ between the fitted values $Y_{fit}$ and calculated numerical results $Y$ has been defined as   

\begin{equation}
\Delta Y = \frac{|Y-Y_{fit}|}{Y_{fit}}
\label{L-tild-2a}
\end{equation}
\noindent 
which provides the fractional difference between the numerical results and the corresponding analytic fits.

\begin{table*}[htbp]
\vspace{0.0cm}
\centering
\caption{\label{tab:table1} The variables and fitted coefficients.}
\begin{tabular}{ccccccc}
\hline
\hline

$Y$&$X$&a$_0$&a$_1$&a$_2$&a$_3$&a$_4$ \\ \hline

${\rm ln}Q$&${\rm ln}M$&
0.232466E+01&-0.166851E+01&-0.709940E+00&-0.159270E+00&0.510313E-01\\
&&&&&&\\

$k_2$&${\rm ln}M$&
0.970581E-01&-0.214968E-01&-0.773239E-01&-0.224548E-01&-0.390682E-03\\
&&&&&&\\

${\rm ln}\Lambda$&${\rm ln}M$& 
0.803693E+01&-0.539839E+01&-0.234640E+01&-0.106376E+01&0.729592E-02\\
&&&&&&\\

${\rm ln}I$&${\rm ln}\Lambda$& 
0.158383E+01&0.113245E+00&0.183125E-01&-0.788120E-03&0.331898E-04\\
&&&&&&\\

$C$&${\rm ln}\Lambda$&
0.360812E+00&-0.357859E-01&0.315680E-03&0.615497E-04&-0.162488E-05\\
&&&&&&\\

${\rm ln}I$&${\rm ln}Q$&
0.133849E+01&0.120787E+01&-0.479594E+00&0.174713E+00&-0.139314E-01\\
&&&&&&\\

${\rm ln}Q$&${\rm ln}\Lambda$&
0.184514E+00&0.107931E+00&0.435450E-01&-0.361704E-02&0.942312E-04\\
&&&&&&\\

$C$&${\rm ln}Q$&
0.354018E+00&-0.117589E+00&0.197844E-02&0.261095E-02&-0.217161E-03\\

\hline
\hline 
\end{tabular} 
\vspace{0.0cm}
\end{table*}
\noindent 
    
    The variation of dimensionless quadrupole moment ($Q$) with NS mass ($M$) obtained for APR, BSk22, BSk24 and BSk26 EoSs are displayed in Fig.-\ref{fig2}. The fitting curve (black dashed), given by Eq.(\ref{L-tild-2}), for the universal relation is also shown. In Figs.-\ref{fig3} and \ref{fig4}, respectively, variations of Love number ($k_2$) and the dimensionless tidal deformability ($\Lambda$) with mass ($M$) are displayed for APR, BSk22, BSk24 and BSk26 EoSs. The fitting curves (black dashed), given by Eq.(\ref{L-tild-2}), for the universal relation for these are also shown. The shaded regions represent the HESS J1731-347 remnant \cite{Dor22}, the GW170817 event \cite{Abbott19}, PSR J1614-2230 \cite{Arz18}, PSR J0348+0432 \cite{Ant13}, PSR J0740+6620 \cite{Cro20}, and PSR J0952-0607 \cite{Rom22} pulsar observations for the possible maximum mass.
    
    The moment of inertia $I_{45}$ in units of 10$^{45}$ g cm$^2$ versus mass $M$ for APR, BSk22, BSk24 and BSk26 EoSs are plotted in Fig.-\ref{fig5}. In Fig.-\ref{fig6} the plots of compactness ($C$) against dimensionless moment of inertia ($I$) for APR, BSk22, BSk24 and BSk26 EoSs corresponding to maximum masses are shown. In Fig.-\ref{fig7} the variation of dimensionless moment of inertia ($I$) with mass ($M$) for APR, BSk22, BSk24 and BSk26 EoSs are shown.

    In Figs.-\ref{fig8},\ref{fig9},\ref{fig10},\ref{fig11},\ref{fig12},\ref{fig13},\ref{fig14},\ref{fig15},\ref{fig16},\ref{fig17}, respectively, the plots of the dimensionless moment of inertia ($I$) versus dimensionless tidal deformability ($\Lambda$), compactness ($C$) versus the dimensionless tidal deformability ($\Lambda$), the dimensionless moment of inertia ($I$) versus  dimensionless quadrupole moment ($Q$), dimensionless quadrupole moment ($Q$) versus  the dimensionless tidal deformability ($\Lambda$) and compactness ($C$) versus dimensionless quadrupole moment ($Q$) are shown for APR, BSk22, BSk24 and BSk26 EoSs corresponding to maximum masses along with respective relative fractional errors between the fitting curve and results of numerical calculations. The fitting curves (black dashed), given by Eq.(\ref{L-tild-2}), for the universal relation for these are also shown. It may be discerned that the universal relations strongly emerge except in cases of variations of $Q$, $k_2$ and $\Lambda$ with NS mass $M$ where the universal relations do not crop up so strongly. 
       
\noindent
\subsection{Chirp Mass} 
\label{Subsection 4b}

    The chirp mass ${\rm M_c}$ of a binary NSs system, having constituent masses $\mathcal{M}_1$ and $\mathcal{M}_2$, is an reliably measurable quantity by the ground-based detectors \cite{Abbott-gw170817} for GWs. It is defined in terms of constituent masses as
    
\begin{equation}
{\rm M_c}=\frac{(\mathcal{M}_1\mathcal{M}_2)^{3/5}}{(\mathcal{M}_1+\mathcal{M}_2)^{1/5}}=\mathcal{M}_1\frac{q^{3/5}}{(1+q)^{1/5}},
\label{chirpmass}
\end{equation}
where by convention $\mathcal{M}_1 \geq \mathcal{M}_2$ has been chosen so that the binary mass ratio $q=\mathcal{M}_2/\mathcal{M}_1$ lies within the range $0 < q\leq1$.

    Another reliably measurable quantity is the dimensionless effective tidal deformability $\Lambda^*$ which is defined as \cite{Abbott-gw170817}
\begin{equation}
\Lambda^*=\frac{16}{13}\frac{(\mathcal{M}_1+12\mathcal{M}_2)\mathcal{M}_1^4\Lambda_1 + (\mathcal{M}_2+12\mathcal{M}_1)\mathcal{M}_2^4\Lambda_2}{(\mathcal{M}_1 + \mathcal{M}_2)^5},
\label{L-tild-1}
\end{equation}
such that $\Lambda^*=\Lambda_1=\Lambda_2$ when $\mathcal{M}_1 = \mathcal{M}_2$ 
where $\Lambda_{1}$ and $\Lambda_{2}$ are the dimensionless tidal deformabilities \cite{Abbott-gw170817} for the corresponding masses. The variations of chirp mass (${\rm M_c}$) as a function of effective tidal deformability ($\Lambda^*$) have been shown for binary mass ratios of $q=0.7,0.9,0.99,1.0$, respectively, in Figs.-\ref{fig18},\ref{fig19},\ref{fig20},\ref{fig21}.

\noindent
\section{Summary and conclusions}
\label{sec:Section 5}

    Our objective in the present work is to explore the properties of neutron stars such as mass ($M$), radius ($R$), compactness ($C$), Love number ($k_2$), dimensionless tidal deformability ($\Lambda$), dimensionless quadrupole moment($Q$) and the dimensionless moment of inertia ($I$) within a perturbative approach using canonical (APR) and Brussels-Montreal Skyrme (BSk22, BSk24, BSk26) EoSs describing hadronic matter of neutron stars. Various relations among these calculated quantities have been studied. Most of these relations have been found not to depend sensitively on the structure of neutron star interior implying universality. According to this concept of universality, measurements of one quantity that occurs in a universal relation inevitably provide information about the others, even though those others might not be observable. These can be used to test General Relativity independently of nuclear structure, quantify spin in binary inspirals by breaking degeneracies in gravitational wave detection and assess the deformability of compact stars by moment of inertia measurements.

    Exciting applications in fundamental physics, GW theory and astrophysics can be made possible by the universal relations. Numerous follow-up investigations can be possible by the examining the universal relations provided here. One could find out, for instance, if these relations hold for compact stars that rotate rapidly, have significant internal magnetic fields and anisotropic pressure. Perhaps the most severe approximation is the slow-rotation approximation which is appropriate only for NSs with periods higher than or equal to one millisecond. It may be anticipated that the EoS universality identified here would withstand even for rapid rotation, but with distinct universal relations for stars with different spin periods. But this variation should be limited within 10$\%$ \cite{Benhar2005,Berti2004,Berti2005,Pappas2012,Yagi2013,Yagi2015}. To allow for fast spinning NSs, one possible expansion of this work would be to improve the universal relations.

\begin{acknowledgements}

    D. N. Basu acknowledges support from Science and Engineering Research Board, Department of Science and Technology, Government of India, through Grant No. CRG/2021/007333.

\end{acknowledgements}		


\end{document}